\documentstyle[mprocl]{article}

\input{psfig}
\input{epsfig.sty}
\def\nn{\noindent}
\def\ie{{\it i.e.}}

\def\etc{{\it etc}}
\def\etal{{\it et al.}}
\def\be{\begin{equation}}
\def\ee{\end{equation}}
\def\bea{\begin{eqnarray}}
\def\eea{\end{eqnarray}}

\begin{document}

\rightline{\vbox{\halign{&#\hfil\cr
&SLAC-PUB-7663\cr
&October 1997\cr}}}
\vspace{0.8in}

\title{{$Z'$ INVESTIGATIONS AT FUTURE LEPTON COLLIDERS}
\footnote{To appear in the {\it Proceedings of the $2^{nd}$ International 
Workshop on $e^-e^-$ Interactions at TeV Energies}, Santa Cruz, CA, 
22-24 September 1997}
}

\author{ {T.G. RIZZO}
\footnote{Work supported by the Department of Energy, 
Contract DE-AC03-76SF00515}
}

\address{Stanford Linear Accelerator Center,\\
Stanford University, Stanford, CA 94309, USA\\
E-mail: rizzo@slacvx.slac.stanford.edu}

\maketitle\abstracts{In this talk I summarize the capability of future 
lepton colliders to indirectly discover a new $Z'$ and to determine its 
couplings to the fermions of the Standard Model. The physics associated with 
sitting on the $Z'$-pole is also briefly discussed. This analysis is based on 
the results presented in the Snowmass 1996 New Gauge Boson Working Group 
report.}

\section{Introduction}

One of the primary goals of present and future collider searches is to 
ascertain the gauge group which describes the electroweak and strong 
interactions. There are many scenarios in the literature wherein the Standard 
Model(SM) gauge group is augmented by at least an additional $U(1)$ factor, 
implying the existence {\cite {steve}} of a new neutral gauge boson, $Z'$. In 
such scenarios the apparent success of the SM at low energies {\cite {rev}} 
is essentially due to decoupling arguments associated with the observation 
that new gauge bosons must most likely be an order of magnitude heavier than 
their SM counterparts. Indeed collider searches {\cite {tev,lep}} indicate 
that a canonical $Z'$ with couplings to both quarks and leptons is probably 
more massive than about 600 GeV. (The reader should, however, remember the 
caveats associated with such strong statements since the number and type of 
$Z'$ models in the literature is quite enormous.) If true, this implies that 
that a $Z'$ will be beyond the direct search reach of a first generation 
lepton collider whose center of mass energy is expected to be 500 GeV or less.

Extended Gauge Models(EGMs) can be divided into two very broad classes 
depending upon whether or not they originate from a GUT group, such 
as $SO(10)$ or 
$E_6$. {\it Generally}, the new gauge bosons from GUT-inspired scenarios have 
generation-independent couplings (in the same sense as the $W$ and $Z$ of the 
SM), whereas this need not be true for non-unifiable models. Also, 
{\it generally}, the extension of the SM group structure induces additional 
anomalies which cannot be cancelled by using the conventional SM fermions 
alone. This implies the almost all EGMs also contain additional exotic matter 
particles, such as leptoquarks, with masses comparable to those of the new 
gauge bosons themselves.
The search reach at a collider for new gauge bosons is somewhat model 
dependent due to the rather large variations in their 
couplings to the SM fermions which are 
present in extended gauge theories currently on the market. 
This implies that any overview of the subject is necessarily incomplete. 
Hence, we will be forced to limit ourselves to a few representative models. 
In what follows, we chose as examples the set of models recently discussed by 
Cvetic and Godfrey {\cite {steve}} so that we need to say very little here 
about the details of the coupling structure of each scenario. To be specific we 
consider ({\it i}) the $E_6$ effective rank-5 model(ER5M), which predicts a 
$Z'$ whose couplings depend on a single parameter 
$-\pi/2 \leq \theta \leq \pi/2$ (with models $\psi$, $\chi$, $I$, and $\eta$ 
denoting specific $\theta$ values); ({\it ii}) the Sequential Standard 
Model(SSM) 
wherein the new $W'$ and $Z'$ are just heavy versions of the SM particles (of 
course, this is not a true model in the strict sense but is commonly used as a 
guide by experimenters); ({\it iii}) the Un-unified Model(UUM), based on the 
group $SU(2)_\ell \times SU(2)_q \times U(1)_Y$, which has a 
single free parameter $0.24 \leq s_\phi \leq 1$; 
({\it iv}) the Left-Right Symmetric Model(LRM), based on the group 
$SU(2)_L \times SU(2)_R \times U(1)_{B-L}$, 
which also has a free parameter $\kappa=g_R/g_L$ of order unity which is just 
the ratio of the gauge couplings 
and, lastly, ({\it v}) the Alternative Left-Right Model(ALRM), based on the 
same extended group as the LRM but now arising from 
$E_6$, wherein the fermion assignments are modified in comparison to the LRM 
due to an ambiguity in how they are embedded in the {\bf 27} representation.

\vspace*{-0.5cm}
\nn
\begin{figure}[htbp]
\centerline{
\psfig{figure=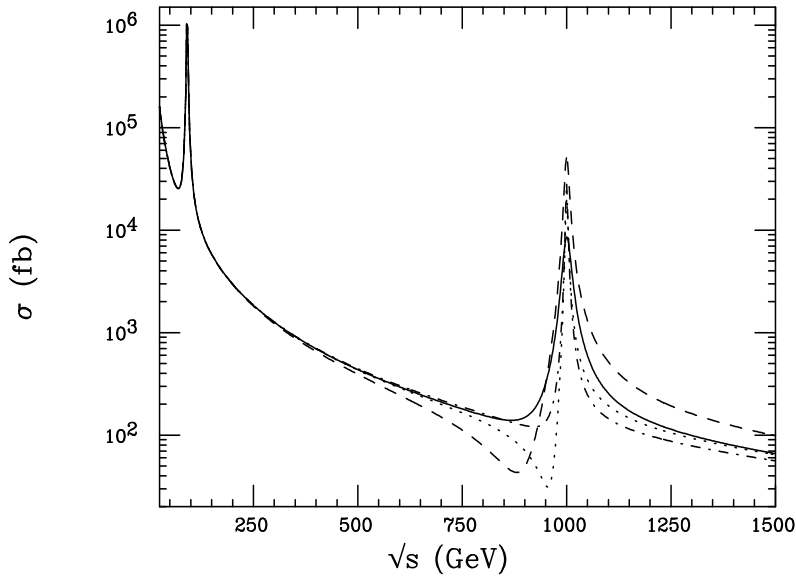,height=9.5cm,width=12cm,angle=-90}}
\vspace*{-0.9cm}
\caption{Cross section for $e^+e^-\to \mu^+\mu^-$ in the presence of a 1 TeV 
$Z'$ that couples only to SM fermions. The solid(dash-dotted, dashed, dotted) 
curve corresponds to the LRM with $\kappa=1$($\psi$, $\chi$, $\eta$). ISR has 
not been included.}
\label{eemumu}
\end{figure}
\vspace*{0.4mm}

\section{On The $Z'$ Peak}

It is quite possible that the LHC may find a $Z'$ in the TeV region before the 
NLC turns on. (As will be seen below, the LHC reach for a canonical $Z'$ 
is 4-5 TeV.) In fact, several arguments suggest, at least in some 
string-motivated SUSY models, that the $Z'$ mass cannot be far above 
1 TeV {\cite {cl}} as it is naturally linked to the scale of electroweak 
symmetry breaking. As was discussed by Cvetic and Godfrey {\cite {steve}} and 
in the more recent analyses presented at Snowmass 1996 {\cite {sno}}, it will 
not be easy for the LHC to uniquely determine the couplings of a $Z'$ if its 
mass is too far above 1 TeV due to a lack of robust observables. Even for 
masses as low as 1 TeV it remains unclear just how well a real LHC detector 
can do in this regard and detailed studies have yet to be performed. Of 
course if we are lucky to have a $\sim 1$ TeV $Z'$ this will be an ideal 
opportunity for a lepton collider although this possibility has 
gotten little attention. Fig.\ref{eemumu} shows the 
cross section for $e^+e^- \to \mu^+\mu^-$ in the presence of a 1 TeV $Z'$ for 
several different EGMs. We see that with an integrated luminosity of 
100 $fb^{-1}$, hundreds of thousands of $\mu$-pair events will be collected 
on the peak even after ISR is accounted for. This suggests that in this case 
the NLC will essentially repeat the analyses of SLC/LEP to determine the $Z'$ 
couplings to the SM fermions, as well as to other exotic final states which are 
kinematically accessible.

\vspace*{-0.5cm}
\nn
\begin{figure}[htbp]
\centerline{
\psfig{figure=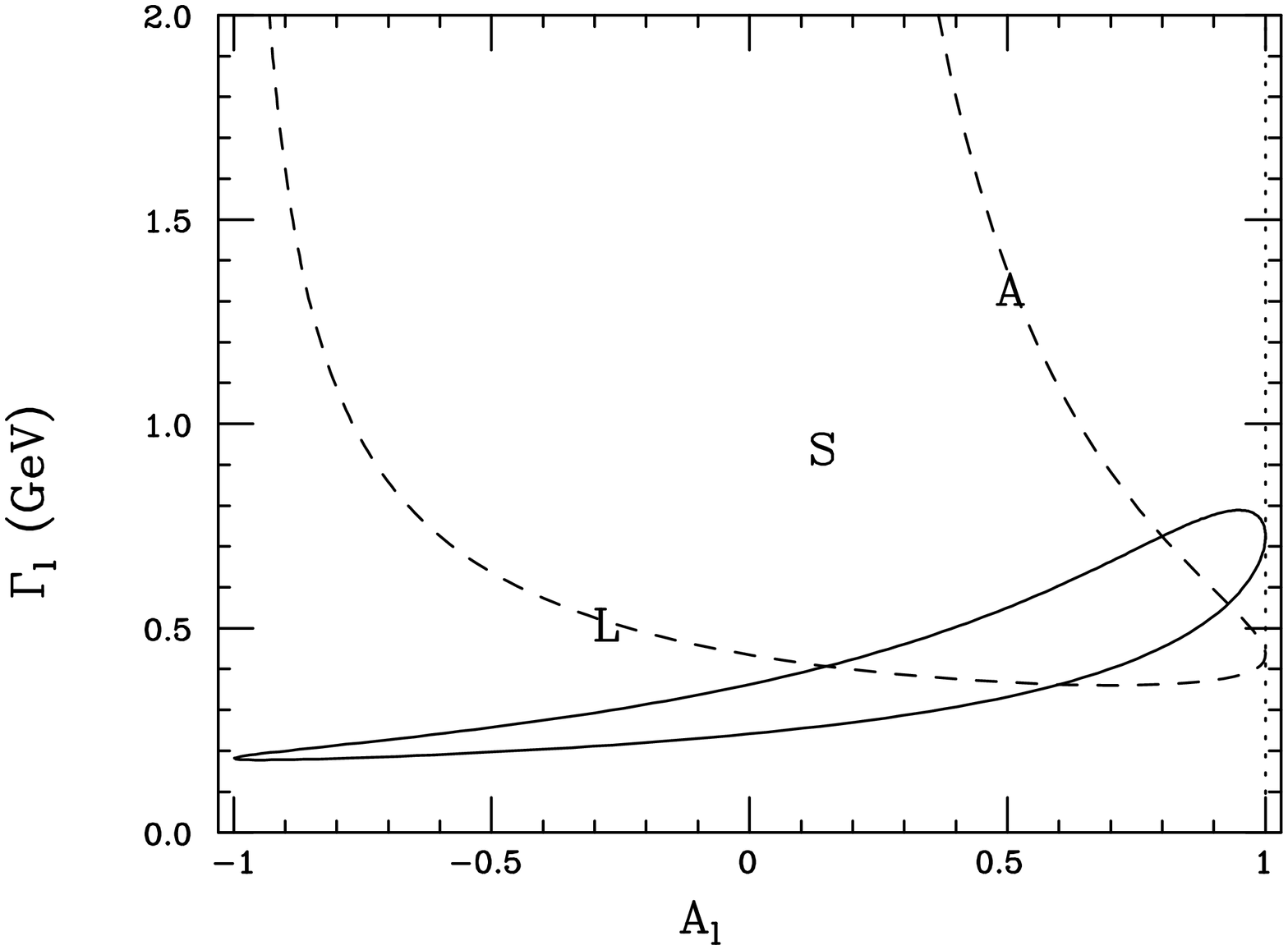,height=6.9cm,width=6.9cm,angle=-90}
\hspace*{-5mm}
\psfig{figure=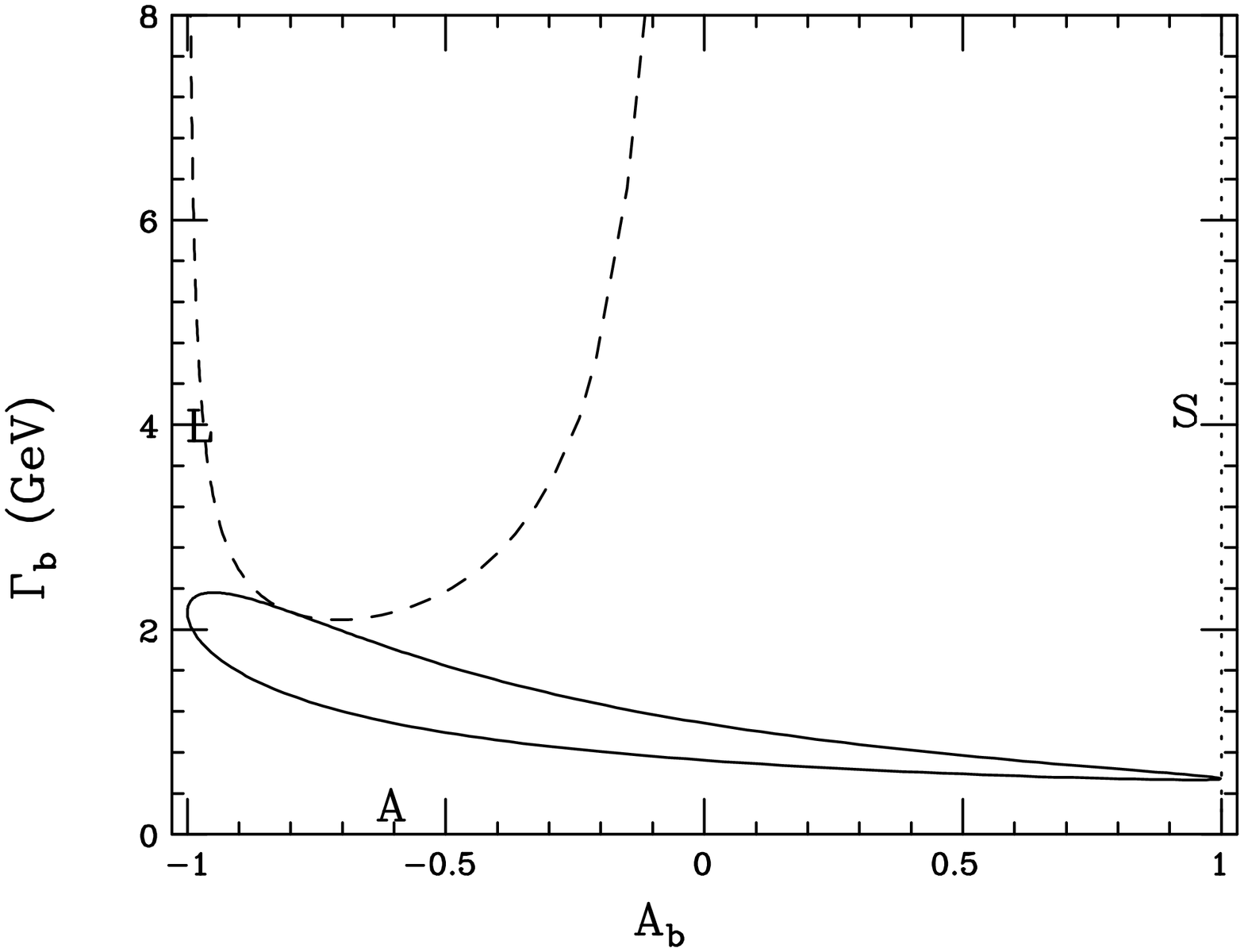,height=6.9cm,width=6.9cm,angle=-90}}
\vspace*{-0.75cm}
\centerline{
\psfig{figure=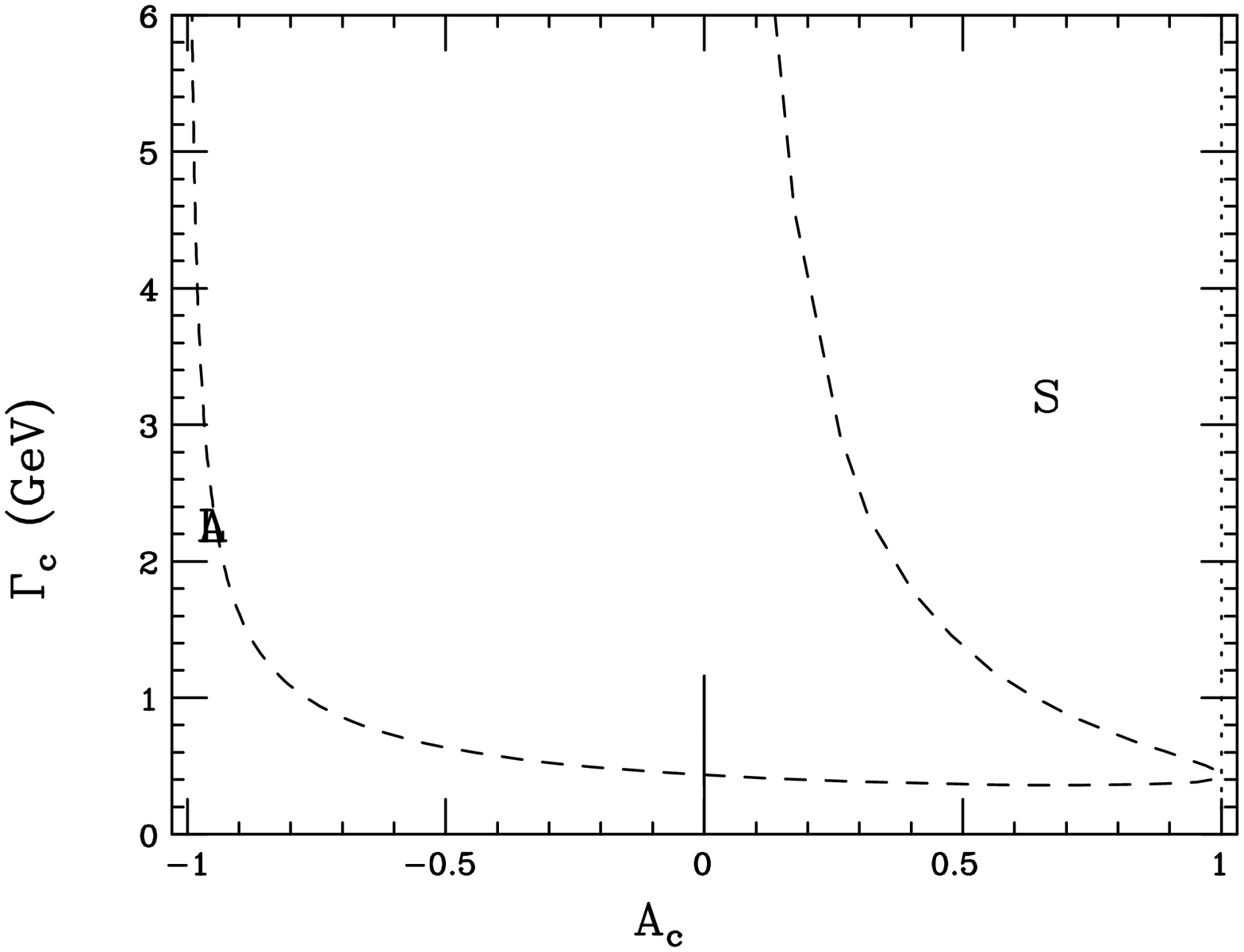,height=6.9cm,width=6.9cm,angle=-90}}
\vspace*{-0.3cm}
\caption{Partial decay widths versus the corresponding asymmetry parameters 
for the $\ell$, $b$ and $c$ final states in $Z'$ decays in $E_6$ 
models(solid), LRM(dashed), UUM(dash-dotted) as well as for the ALRM(A), the 
SSM(S) and the LRM with $\kappa=1$(L). The $Z'$ is assumed to have a mass of 
1 TeV. $A_f=2v_fa_f/(v_f^2+a_f^2)$, where $v_f,a_f$ are the $Z'$ couplings to 
fermion $f$.}
\label{tom1}
\end{figure}

Fig.\ref{tom1} shows that by using leptonic, $b$, and $c(t)$ final states it 
will be quite trivial to distinguish among the more popular models. We note 
that the anticipated size of the errors associated with any of these 
observables, given the large statistics available and the expected performance 
of an NLC detector, will be comparable to the thickness of one of the lines on 
these plots! We further note that none of the chosen observables 
depend upon the possibility that the $Z'$ may have decays to other than SM 
particles. A $Z'$ in the TeV range will provide an excellent object for study 
at a lepton collider.

\section{Indirect Searches}

\begin{figure}[htbp]
\centerline{
\psfig{figure=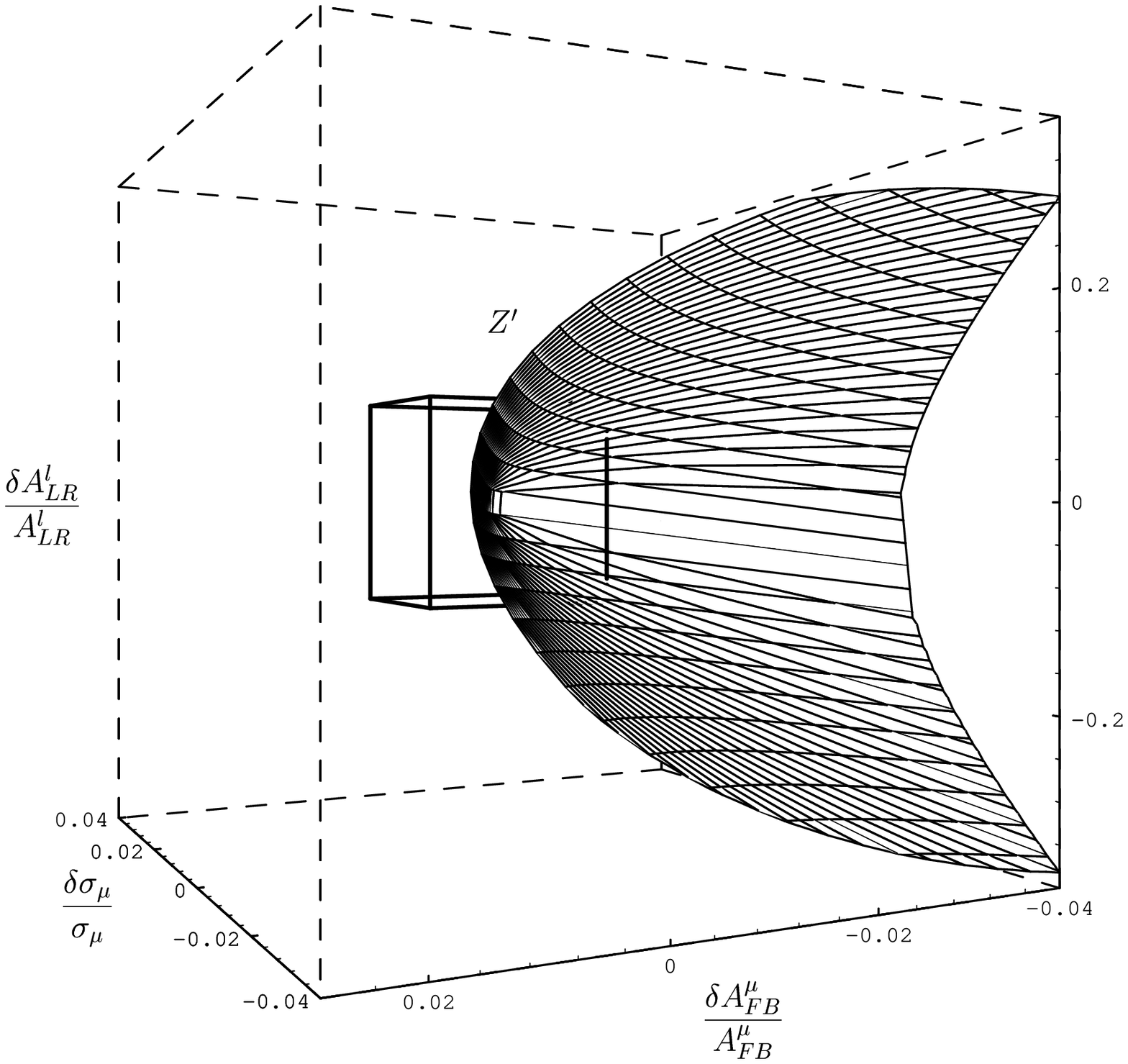,height=7.7cm,width=6.9cm,angle=0}
\hspace*{-7mm}
\psfig{figure=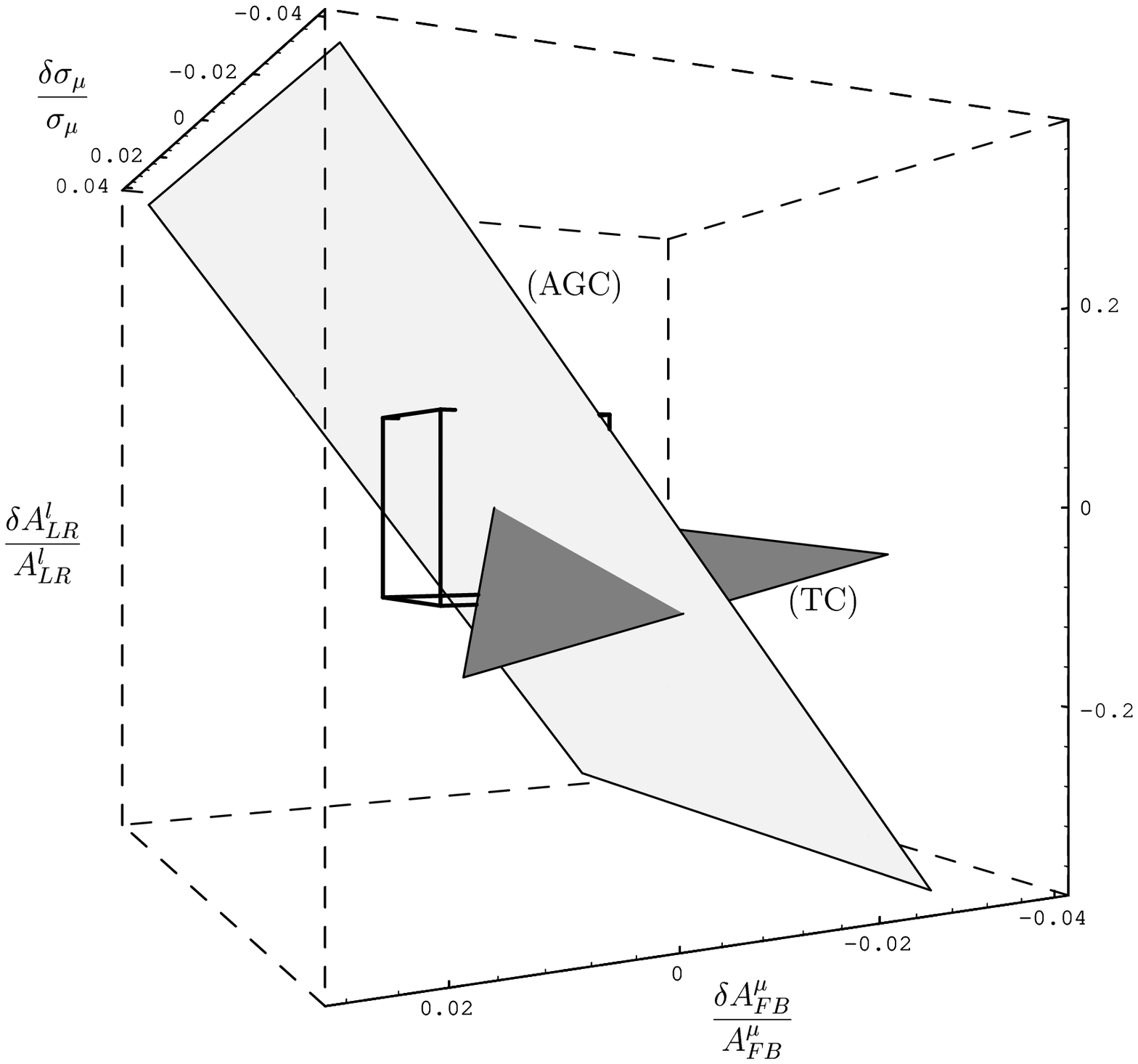,height=7.7cm,width=6.9cm,angle=0}}
\vspace*{-0.9cm}
\caption{Relative shifts in $\sigma_l$, $A_{FB}^l$ and $A_{LR}^l$ due to a 
general $Z'$(left) compared to models with anomalous gauge couplings(light 
grey) or Technicolor models(dark grey) at a 500 GeV NLC taken from the 
analysis of Renard and Verzegnassi $^7$.}
\label{euro}
\end{figure}

It is more than likely that we will be unlucky and a $Z'$ will be too massive 
to be produced directly 
at the first generation of new lepton colliders. Thus searches at such 
machines will be indirect and will consist of 
looking for deviations in the predictions of the SM in as many observables as 
possible. (That such deviations are observable below the $Z'$ peak is obvious 
from Fig.\ref{eemumu}.) 
As is well known, in general analyses of this kind 
the following standard set of observables are employed: 
$\sigma_{f}$, $A_{FB}^{f}$, $A_{LR}^{f}$, $A_{pol}^{FB}(f)$ 
where $f$ labels the fermion in the final state and, special to the case of 
the tau, $<P_\tau>$ and $P_\tau^{FB}$. Note 
that beam polarization plays an important role in this list of observables, 
essentially doubling its length. Layssac \etal ~and, more recently, Renard and 
Verzegnassi {\cite {lay}} have shown that the deviations in the 
leptonic observables due to the existence of a $Z'$ are rather unique and 
differ from other new virtual effects as shown in Fig.\ref{euro}.

\vspace*{-0.5cm}
\nn
\vspace*{0.1mm}

\begin{figure}[htbp]
\centerline{
\psfig{figure=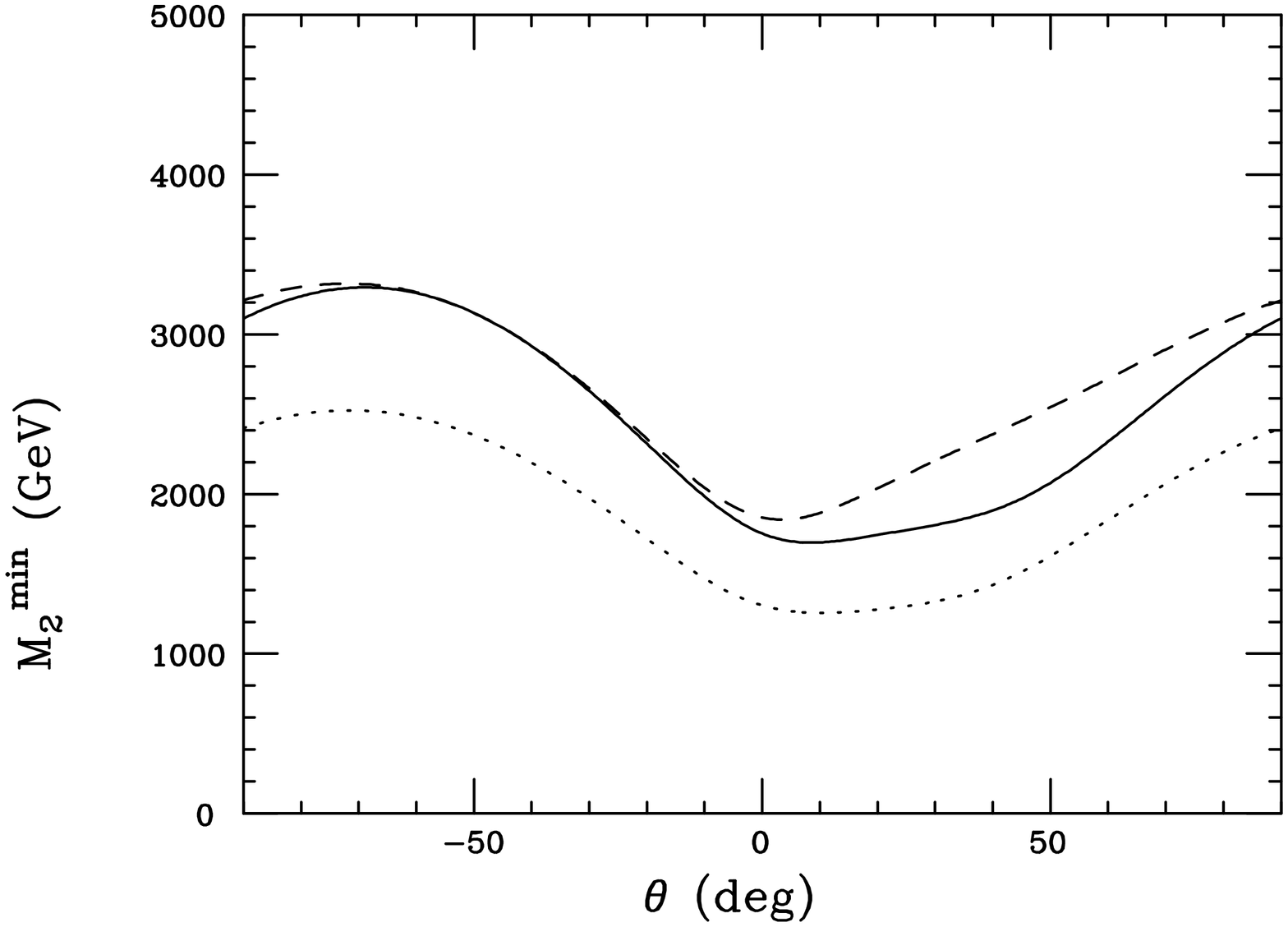,height=7.5cm,width=6.9cm,angle=-90}
\hspace*{-5mm}
\psfig{figure=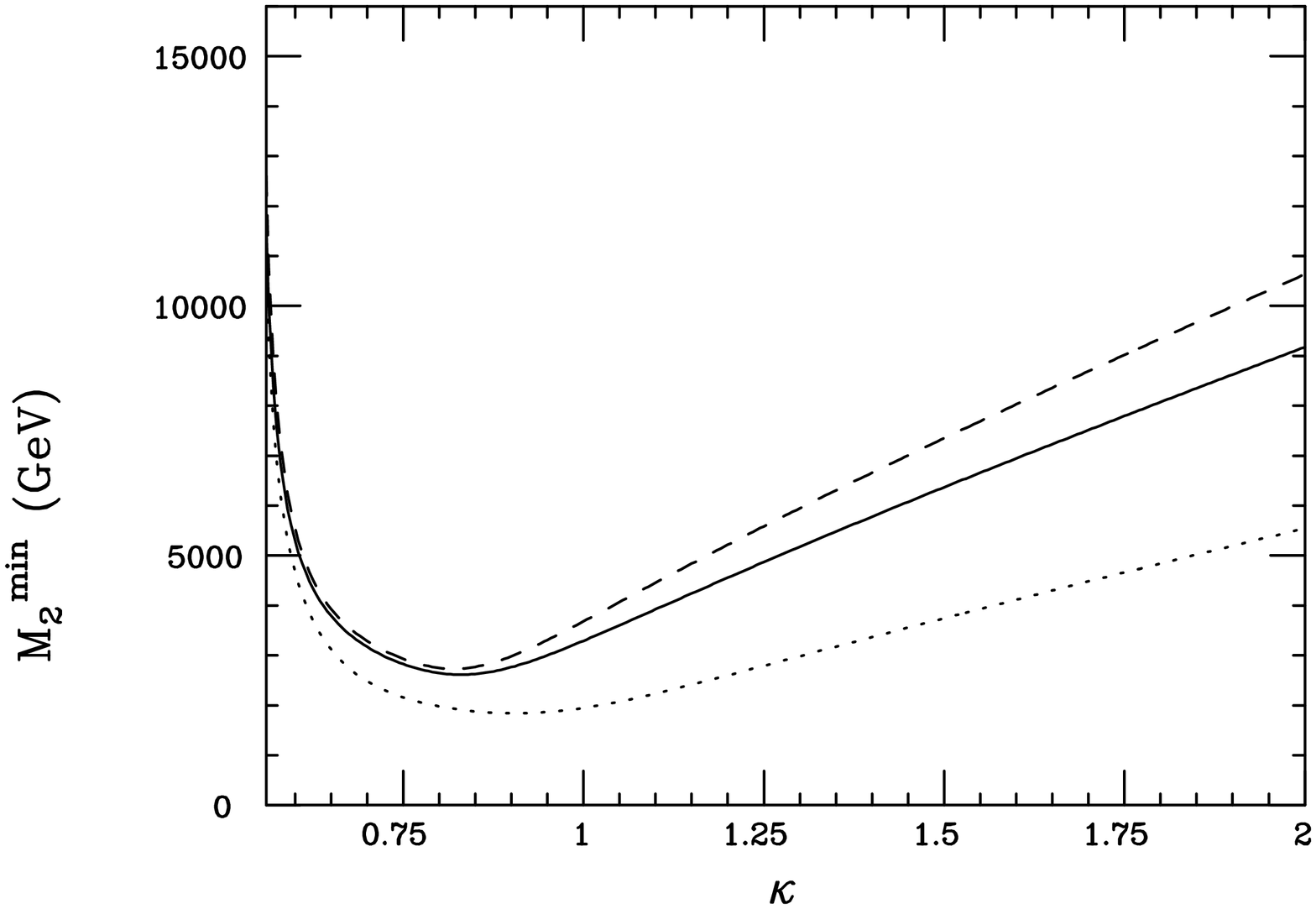,height=7.5cm,width=6.9cm,angle=-90}}
\vspace*{-0.6cm}
\caption{Indirect $Z'$ search reaches at the 500 GeV NLC for $E_6$ models as 
a function of $\theta$ and the LRM as a function of $\kappa$ including initial 
state radiation. The dotted(solid, dashed) curve corresponds to the values 
obtained using leptonic(leptonic plus $b-$quark, all) observables. A luminosity 
of 50 $fb^{-1}$ has been assumed.}
\label{fignlc}
\end{figure}

Fig.\ref{fignlc} from {\cite {sno}} shows the resulting search reach from this 
kind of analysis for the 500 GeV NLC assuming a $Z'$ in either the ER5M 
or the LRM employing observables from 
$l,b,c$ and $t$ final states including systematic effects, detector cuts, 
ISR, \etc. $90\%$ beam polarization was assumed. Here we can directly see the 
importance of 
employing as many final states as possible. Table~\ref{tabnlc} shows the 
corresponding results for several lepton colliders, \ie, $e^+e^-$ colliders 
with $\sqrt s$= 0.5, 1, 1.5 and 5 TeV as well as a 4 TeV muon collider, 
in comparison to the direct 
LHC search reaches from {\cite {sno}}. A very interesting analysis describing 
the scaling behaviour of $Z'$ reaches at lepton colliders with both energy 
and luminosity was quite recently performed by Leike {\cite {leike}} to which 
we refer the interested reader.

\begin{figure}[htbp]
\leavevmode
\centerline{
\input{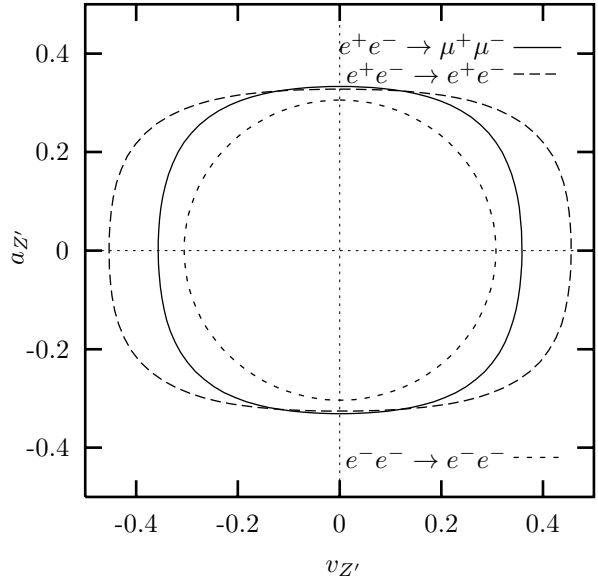}}
\vspace*{-0.3cm}
\caption{Contours of observability at 95$\%$ CL for the reduced $Z'$ couplings 
including the effects of ISR, polarization and luminosity uncertainties, as 
well as the angular resolution of the detector. These results are for a 500 
GeV NLC with $P=90\%$ with a luminosity of 50(25) $fb^{-1}$ in the 
$e^+e^-(e^-e^-)$ mode.}
\label{cuypersfig}
\end{figure}

As has been discussed by Cuypers {\cite {cuy}}, $e^-e^-$ collisions offer a 
unique advantage for $Z'$ hunting in comparison to $e^+e^-$ in that greater 
statistics are available and that both beams can be polarized. The 
disadvantage of $e^-e^-$ is the lack of a $Z'$ tail in the $s$-channel. The 
Cuypers analysis {\cite {cuy}} demonstrated that, using leptonic modes only,  
$e^-e^-$ generally has a superior $Z'$ search reach than does $e^+e^-$. The 
same analysis also shows that the {\it ratio} of $e^-e^-$ to $e^+e^-$ search 
reaches was essentially independent of ISR. A comparison of the $Z'$ search 
reaches in the $e^+e^- \to \mu^+\mu^-$, $e^+e^- \to e^+e^-$, and 
$e^-e^- \to e^-e^-$ channels by Cuypers is shown in Fig.\ref{cuypersfig}. 

\vspace*{-0.5cm}
\nn
\vspace*{0.1mm}
%

\begin{table}[htpb]
\caption{Indirect $Z'$ search reaches of lepton colliders in TeV employing 
all observables including the effects of cuts, ISR, \etc. The integrated 
luminosities of the NLC500, NLC1000, NLC1500, NNLC and the Large Muon 
Collider are assumed to be 50, 100, 100, 1000 and 1000 $fb^{-1}$, 
respectively. In the last column we show the corresponding direct search 
reach for the LHC assuming an integrated luminosity of 100 $fb^{-1}$ and $Z'$ 
decays to only SM fermions.}
\centering
\begin{tabular}{lcccccc}
\hline
\hline
Model  & NLC500  &NLC1000 &NLC1500 &NNLC 5 TeV& LMC 4 TeV & LHC \\
\hline
$\chi$  & 3.21 & 5.46 & 8.03 & 23.2 & 18.2 & 4.49\\
$\psi$  & 1.85 & 3.24 & 4.78 & 14.1 & 11.1 & 4.14\\
$\eta$  & 2.34 & 3.95 & 5.79 & 16.6 & 13.0 & 4.20\\
I       & 3.17 & 5.45 & 8.01 & 22.3 & 17.5 & 4.41\\
SSM     & 3.96 & 6.84 & 10.1 & 29.5 & 23.2 & 4.88\\
ALRM    & 3.83 & 6.63 & 9.75 & 28.4 & 22.3 & 5.21\\
LRM     & 3.68 & 6.28 & 9.23 & 25.6 & 20.1 & 4.52\\
UUM     & 4.79 & 8.21 & 12.1 & 34.7 & 27.3 & 4.55\\
\hline
\hline
\end{tabular}
\label{tabnlc}
\end{table}

Table~\ref{cuypers} from {\cite {sno}} shows the ratio of $e^-e^-$ to 
$e^+e^-$ $Z'$ search reaches 
at a 500 GeV collider as different final states are added in the $e^+e^-$ 
case. It confirms the result that if only leptonic observables are employed 
then the $e^-e^-$ reach is superior to $e^+e^-$. However, as soon as 
one adds the additional information from the quark sector, $e^+e^-$ regains 
the lead in terms of $Z'$ mass reach. Combining the leptonic and quark data 
together in the $e^+e^-$ case always results in a small value for the reach 
ratio.

\vspace*{-0.5cm}
\nn
\vspace*{0.4mm}
%

\begin{table}[htbp]
\caption{Ratio of $e^-e^-$ to $e^+e^-$ indirect $Z'$ search reaches at a 500 
GeV NLC with an integrated luminosity of 50 $fb^{-1}$ in either collision 
mode. ISR 
has been ignored. The columns label the set of the final state fermions used in 
the $e^+e^-$ analysis.}
\begin{center}
\label{$e^-e^-$}
\begin{tabular}{lccc}
\hline
\hline
Model  & $\ell$  & $\ell+b$ & $\ell+b,c,t$ \\
\hline
$\chi$  &  1.10 & 0.900 & 0.896 \\
$\psi$  &  1.20 & 0.711 & 0.673 \\
$\eta$  &  1.07 & 0.813 & 0.650 \\
I       &  1.06 & 0.813 & 0.813 \\
SSM     &  1.30 & 0.752 & 0.667 \\
ALRM    &  1.20 & 1.12  & 0.909 \\
LRM     &  1.02 & 0.483 & 0.432 \\
UUM     & 0.891 & 0.645 & 0.496 \\
\hline
\hline
\end{tabular}
\end{center}
\label{cuypers}
\end{table}

\section{Coupling Determinations}

\begin{figure}[htbp]
\leavevmode
\centerline{
\epsfig{file=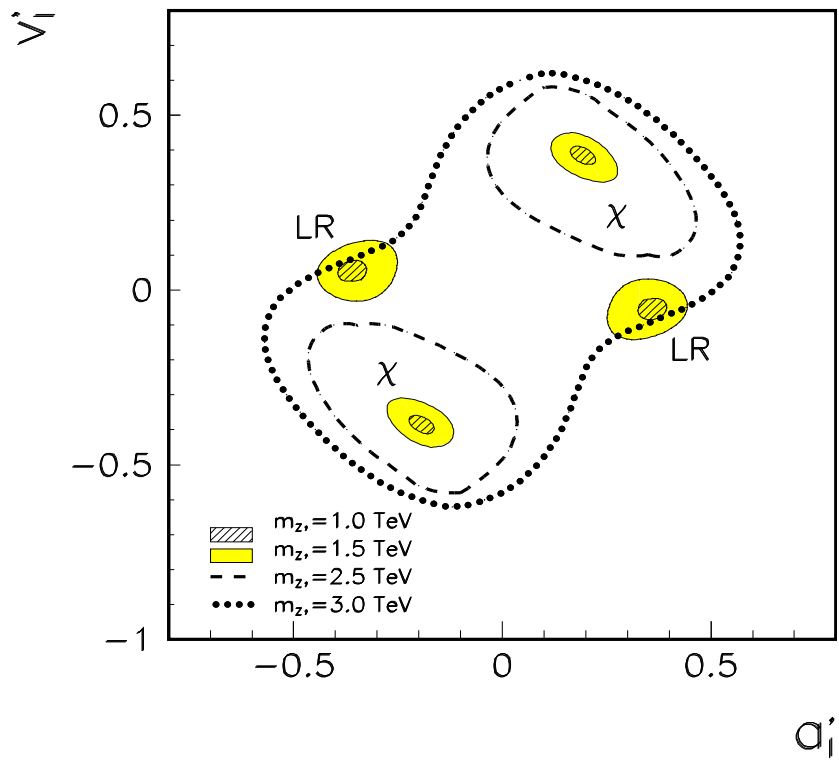,height=10.0cm,width=11cm,clip=}}
\vspace*{-0.3cm}
\caption{$95\%$ CL contours for $v'_l$ and $a'_l$ for a 500 GeV NLC with a 
luminosity of $50fb^{-1}$. The $Z'$ is taken to be in 
the $\chi$ or LRM with a 1(1.5) TeV mass corresponding to the hatched(shaded) 
area. The dashed(dotted) contours are $95\%$ CL limits on the $Z'll$ couplings 
for the $\chi$ case and a mass of 2.5(3) TeV. A beam polarization of 80$\%$ 
has been assumed.}
\label{sabine1}
\end{figure}

If the $Z'$ is not too massive in comparison to $\sqrt s$, then sufficient 
statistical power may be available to not only indirectly see the effects of 
the $Z'$ but also to extract coupling information. 
For example, Riemann {\cite {riemann}} has recently analyzed the 
capability of future $e^+e^-$ colliders operating below the $Z'$ resonance 
to measure the $Z\bar ff$ couplings, where $f=\ell,b,c$. Her analysis 
implicitly assumed that the mass of the $Z'$ was already known(from the LHC) 
and was explicitly used as 
an input into the numerical extraction of couplings. Fig.~\ref{sabine1} shows 
the capability of the NLC running at different energies to measure the 
leptonic couplings of the $Z'$ in the LRM and ER5M $\chi$ as the gauge boson 
mass is varied. It's clear from this analysis that with reasonable 
luminosities the NLC will be able to extract leptonic coupling information for 
$Z'$ masses up to $2-3\sqrt s$. (We recall that the {\it search reach} was 
found to be $6-10\sqrt s$.) Riemann further showed that is was also possible to 
constrain the $c$ and $b$ quark couplings of the $Z'$. As the author correctly 
points out, the size of the systematic errors for the measurements on these 
final states is rather critical to this program. For example, for a 
$Z_\chi$($Z_\psi$) with a 1 TeV mass, the size of the allowed region in the 
$v'_b-a'_b$($v'_c-a'_c$) plane approximately doubles at a 500 GeV NLC with a
luminosity of 50 $fb^{-1}$ if a systematic error of 1(1.5)$\%$ is added to all 
relevant observables. However, as Riemann has shown, the NLC will still be 
able to extract coupling information and distinguish various models using the 
$c,b$ final states.

\vspace*{-0.5cm}
\nn
\vspace*{0.4mm}

A more complex situation arises in the case when the $Z'$ mass is not known 
{\it a priori}. It is clear in this 
circumstance that measurements taken at a single value of $\sqrt s$ will not 
be able to disentangle $Z'$ mass and coupling information. The reason is 
straightforward: to leading order in $s/M_{Z'}^2$, rescaling all of the 
couplings and the value of $Z'$ mass by a common factor would leave all of the 
observed deviations from the SM invariant. In this approximation, the $Z'$ 
exchange appears only as a contact interaction. 
The only potential solution to this problem lies in obtaining data on the 
deviations from the SM predictions at several different values of $\sqrt s$ 
and combining them together in a single fit. 
A first analysis of this kind was performed for Snowmass 1996 {\cite {sno}}, 
in which data from different values of $\sqrt s$ are combined. Only the 
leptonic and $b$-quark couplings to the $Z'$ were considered. For $Z'$ 
masses in the 1.5-2 TeV range which were {\it a priori} unknown, this analysis 
found that combining data taken at 500, 750 and 1000 GeV was sufficient to 
determine the 4 unknown couplings as well as the $Z'$ mass. To insure 
model-independence, the mass and couplings were chosen {\it randomly} and 
{\it anonymously} from rather large ranges.

\vspace*{-0.5cm}
\nn
\begin{figure}[htbp]
\centerline{
\psfig{figure=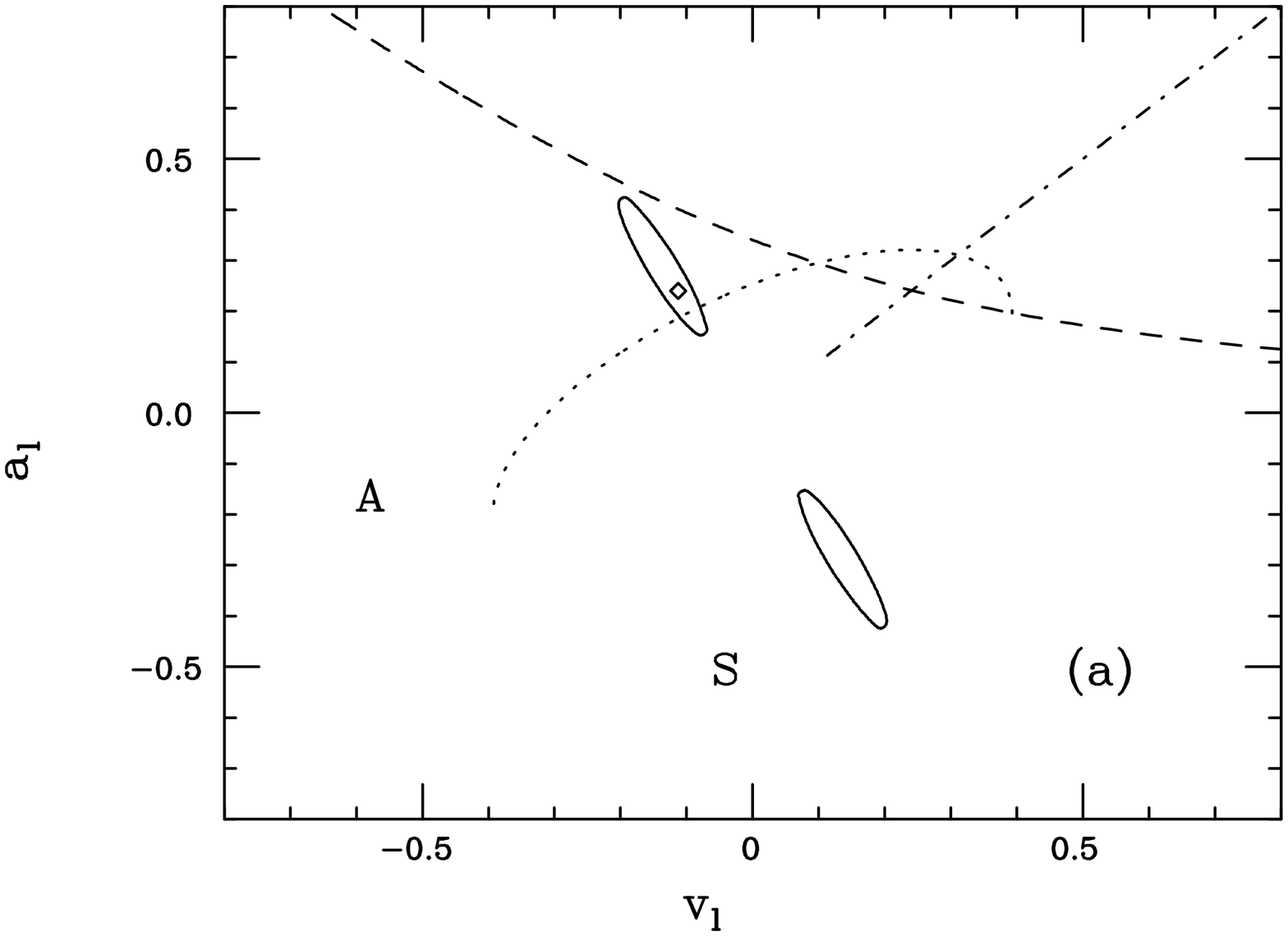,height=6.9cm,width=6.9cm,angle=-90}
\hspace*{-5mm}
\psfig{figure=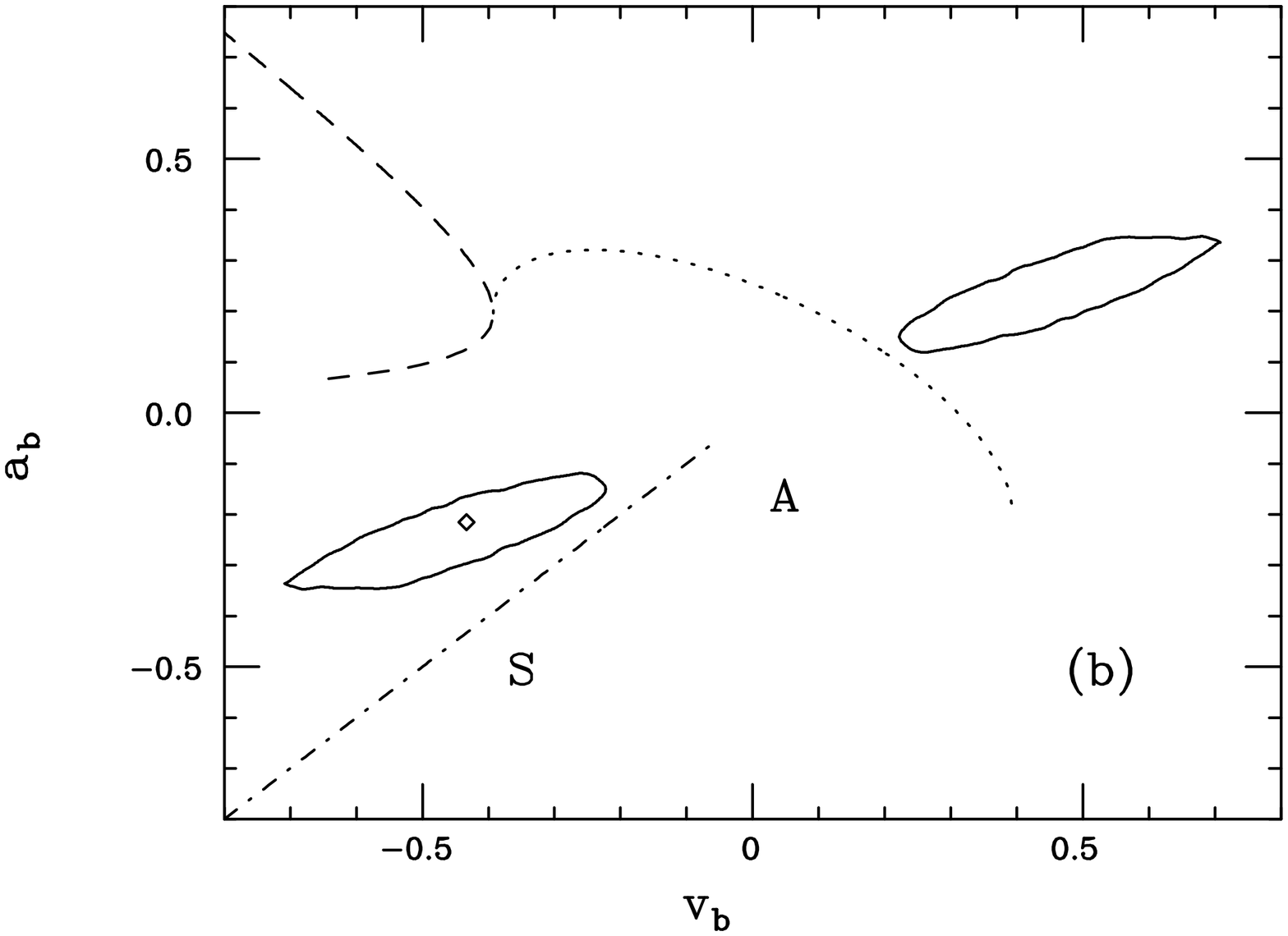,height=6.9cm,width=6.9cm,angle=-90}}
\vspace*{-0.75cm}
\centerline{
\psfig{figure=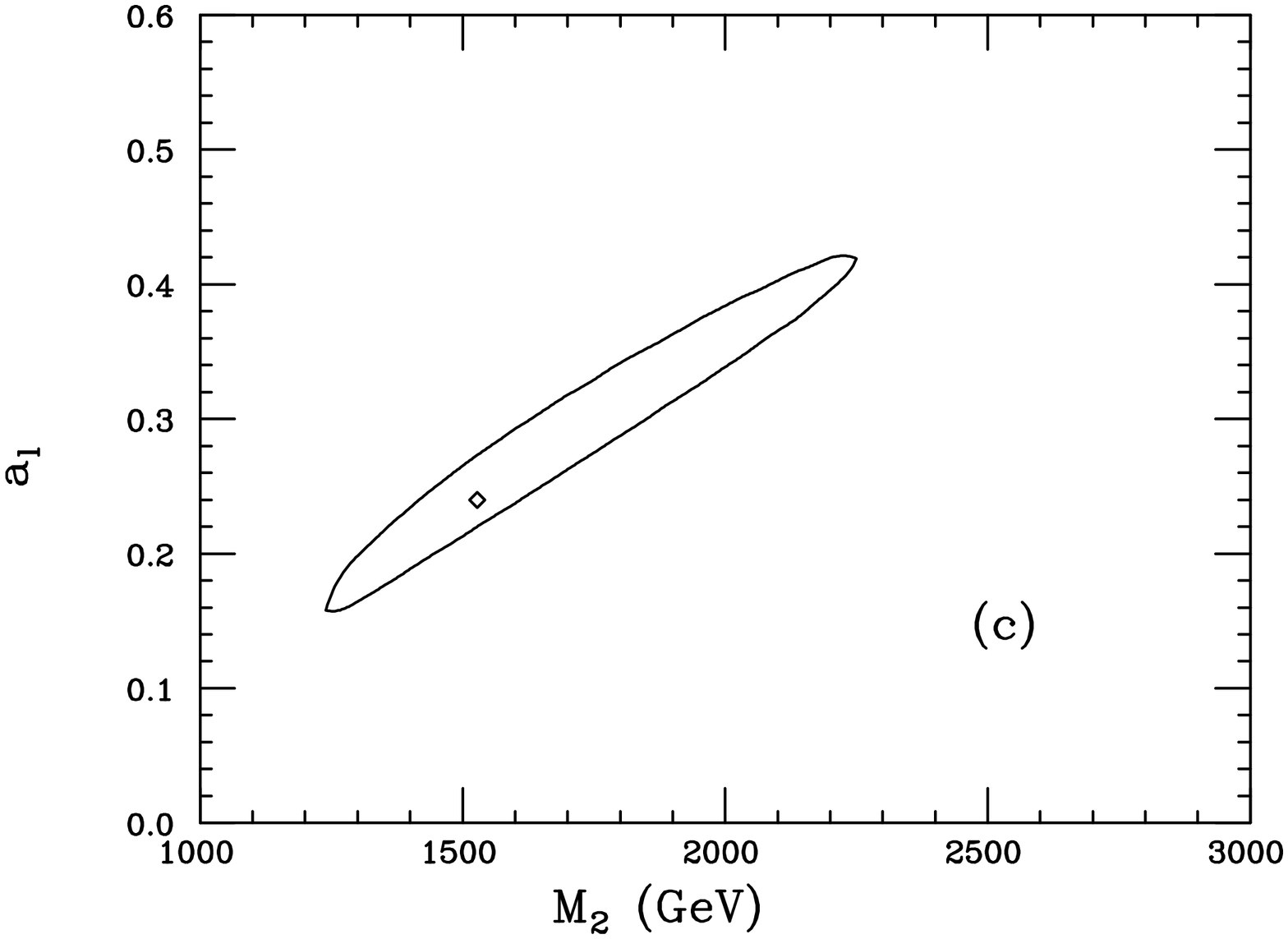,height=6.9cm,width=6.9cm,angle=-90}}
\vspace*{-0.3cm}
\caption{\small $95\%$ CL allowed regions for the extracted values of the 
(a) lepton and (b) $b$-quark couplings 
for a $Z'$ with randomly generated mass and couplings compared with the 
predictions 
of the $E_6$ model(dotted), the Left-Right Model(dashed), and the Un-unified 
Model(dash-dot), 
as well as the Sequential SM and Alternative LR Models(labeled by `S' and `A', 
respectively.) (c) Extracted $Z'$ mass; only the $a_\ell >0$ branch is shown. 
In all cases the diamond represents the corresponding input values. Here we 
seer that the couplings of this $Z'$ do not correspond to those of any of our 
favorite models.}
\label{tom2}
\end{figure}

A sample result of this procedure is shown in Fig.~\ref{tom2}. The three 
figures correspond to two-dimensional projections of the full five dimensional 
($v'_l,a'_l,v'_b,a'_b,M_{Z'}$) $95\%$ CL fit. The following standard set of 
observables were employed: $\sigma_{f}$, $A_{FB}^{f}$, $A_{LR}^{f}$, 
$A_{pol}^{FB}(f)$ where $f=\ell,b$ labels the fermion in the final state and, 
special to the case of the tau, $<P_\tau>$ and $P_\tau^{FB}$. Universality 
amongst the generations was also assumed. While none of the couplings are 
extremely well determined we learn enough to rule out all conventional 
extended gauge models as the origin of this particular $Z'$. Note that 
knowledge of both 
the leptonic and $b-$quarks couplings was required to rule out the case of 
an $E_6$ $Z'$.

\section{Summary and Outlook}

The phenomenology of extended gauge sectors is particularly rich. Analyses have 
evolved in sophistication to the point where detector considerations are 
becoming increasingly important. Many of the problems associated with the 
determination of the couplings of new gauge bosons now have to be faced with 
specific detector capabilities in mind. Although much work has been done, 
there is still a lot of work to be done in the future. Hopefully some of this 
will be completed before new gauge bosons are discovered.

\def\MPL #1 #2 #3 {Mod. Phys. Lett. {\bf#1},\ #2 (#3)}
\def\NPB #1 #2 #3 {Nucl. Phys. {\bf#1},\ #2 (#3)}
\def\PLB #1 #2 #3 {Phys. Lett. {\bf#1},\ #2 (#3)}
\def\PR #1 #2 #3 {Phys. Rep. {\bf#1},\ #2 (#3)}
\def\PRD #1 #2 #3 {Phys. Rev. {\bf#1},\ #2 (#3)}
\def\PRL #1 #2 #3 {Phys. Rev. Lett. {\bf#1},\ #2 (#3)}
\def\RMP #1 #2 #3 {Rev. Mod. Phys. {\bf#1},\ #2 (#3)}
\def\ZPC #1 #2 #3 {Z. Phys. {\bf#1},\ #2 (#3)}
\def\IJMP #1 #2 #3 {Int. J. Mod. Phys. {\bf#1},\ #2 (#3)}

\eject

\end{document}